\documentclass{emulateapj}
\usepackage{graphics}

\slugcomment{Accepted by The Astrophysical Journal}

\newcommand{\chandra}{\textit{Chandra }}

\newcommand{\xmm}{\textit{XMM }}
\newcommand{\msh}{MSH~15--5\textit{6}}

\begin{document}


\shortauthors{TEMIM ET AL.}

\shorttitle{HIGH-ENERGY EMISSION FROM MSH~15--5\textit{6}} 

\title{HIGH-ENERGY EMISSION FROM THE COMPOSITE SUPERNOVA REMNANT MSH~15--5\textit{6}}

\author{TEA TEMIM\altaffilmark{1,2}, PATRICK SLANE\altaffilmark{3}, DANIEL CASTRO\altaffilmark{4}, PAUL PLUCINSKY\altaffilmark{3}, JOSEPH GELFAND\altaffilmark{5} AND JOHN R. DICKEL\altaffilmark{6}}

\altaffiltext{1}{Observational Cosmology Lab, Code 665, NASA Goddard Space Flight Center, Greenbelt, MD 20771, USA}
\altaffiltext{2}{Oak Ridge Associated Universities (ORAU), Oak Ridge, TN  37831, USA; tea.temim@nasa.gov}
\altaffiltext{3}{Harvard-Smithsonian Center for Astrophysics, 60 Garden Street, Cambridge, MA 02138, USA}
\altaffiltext{4}{MIT Kavli Institute}
\altaffiltext{5}{New York University Abu Dhabi, UAE}
\altaffiltext{6}{University of New Mexico}


\begin{abstract}

MSH~15--5\textit{6} (G326.3-1.8) is a composite supernova remnant (SNR) that consists of an SNR shell and a displaced pulsar wind nebula (PWN) in the radio. We present \textit{XMM-Newton} and \chandra X-ray observations of the remnant that reveal a compact source at the tip of the radio PWN and complex structures that provide evidence for mixing of the supernova (SN) ejecta with PWN material following a reverse shock interaction. The X-ray spectra are well fitted by a non-thermal power-law model whose photon index steepens with distance from the presumed pulsar, and a thermal component with an average temperature of 0.55 keV. The enhanced abundances of silicon and sulfur in some regions, and the similar temperature and ionization timescale, suggest that much of the X-ray emission can be attributed to SN ejecta that have either been heated by the reverse shock or swept up by the PWN. We find one region with a lower temperature of 0.3 keV that appears to be in ionization equilibrium. Assuming the Sedov model, we derive a number of SNR properties, including an age of 16,500 yr. Modeling of the $\gamma$-ray emission detected by \textit{Fermi} shows that the emission may originate from the reverse shock-crushed PWN.

\end{abstract}


\section{INTRODUCTION} \label{intro}

Composite supernova remnants (SNRs) are those consisting of a central pulsar that produces a wind of synchrotron-emitting relativistic particles, and a supernova (SN) blast wave that expands into the surrounding interstellar medium (ISM). The interactions between the SNR and the evolving pulsar wind nebula (PWN) can reveal a wealth of information about the SNR properties, the SN ejecta, the central pulsar, and the spectrum of particles injected into the PWN.  At the late stages of a composite SNR's evolution, the SN reverse shock crushes the PWN, resulting in complex filamentary structures and mixing of the PWN material with ejecta gas \citep[e.g.][]{blo01,swa04}. In cases where the reverse shock interacts with the PWN asymmetrically, either due to the pulsar's motion or a density gradient in the ambient ISM, the PWN can be swept away from the pulsar, resulting in a relic PWN \citep[e.g.][]{swa04}. Due to the burn-off of high-energy particles, and the fact that new particles are no longer being injected into the relic PWN, their emission is usually observed at radio wavelengths. As in the case of G327.1-1.1 \citep{tem09}, the pulsar and its newly forming PWN are displaced from the relic nebula. The interaction between the reverse shock and the PWN may also play an important role in the origin of gamma-ray emission in composite SNRs \citep[e.g.][]{sla12}. The increase in the magnetic field due to such an interaction produces an excess of low energy particles that may give rise to gamma-ray emission through inverse Compton scattering.

With its first detection at radio wavelengths, the SNR MSH~15--5\textit{6} (G326.3-1.8) became the prototype for the composite class  of SNRs \citep{mil61,mil79}. A lower limit on its distance of 3.1 kpc has been established by the \ion{H}{1} absorption profile of \citet{gos72}, while \citet{ros96} found a distance of 4.1 kpc through $\rm H\alpha$ velocity measurements, the distance we adopt in this paper. 
Higher-resolution radio observations taken with the Molonglo Observatory Synthesis Telescope (MOST) \citep{whi96} and the Australia Telescope Compact Array (ATCA) showed a symmetric SNR shell, and plerionic component that is displaced from the geometric center of the SNR \citep{gre97,dic00}, reminiscent of a relic nebula that has been disrupted by the reverse shock. The radio emission is non-thermal, with a spectral index of $\alpha = 0.34$ for the non-thermal shell emission, where $S_\nu \propto \nu^{-\alpha}$, and $0.18$ for the PWN component \citep{dic00}. The PWN component is highly polarized, with a luminosity of $\rm L_{10^7-10^{11}Hz}\sim 5\times10^{34}\: erg\:s^{-1}$, making the radio PWN in MSH~15--5\textit{6} the third most luminous after the Crab and G328.4+0.2 \citep{dic00}. A search for the pulsations from the pulsar that powers the nebula have been unsuccessful \citep{kas96}.

Optical $\rm H\alpha$ filaments were observed on the face of the SNR by \citet{ber79}, and appear to correlate spatially with the SNR shell. MSH~15--5\textit{6} was also detected at X-ray wavelengths by ROSAT \citep{kas93} and ASCA \citep{plu98}. The X-ray morphology shows a complete shell that spatially correlates with the radio SNR. The emission appears to be clumpy, with an enhancement in the regions near the PWN. Spectral fits to the ROSAT data showed that the global X-ray emission from 0.1-2.4 keV can be fit with a column density of $\rm N_H=8.9\times10^{21}\:cm^{-2}$ and a thermal model with a temperature of 0.56 keV \citep{kas93}.

In this paper, we report the analysis of the \textit{XMM-Newton} and \chandra observations of MSH~15--5\textit{6} that provide insight into the nature of the X-ray thermal emission and the evolutionary state of the remnant. We also model the {\it Fermi Gamma-ray Space Telescope} Large Area Telescope ({\it Fermi}-LAT) gamma-ray emission that appears to be associated with the SNR and discuss its possible origins.

\begin{figure}
\epsscale{1.1} \plotone{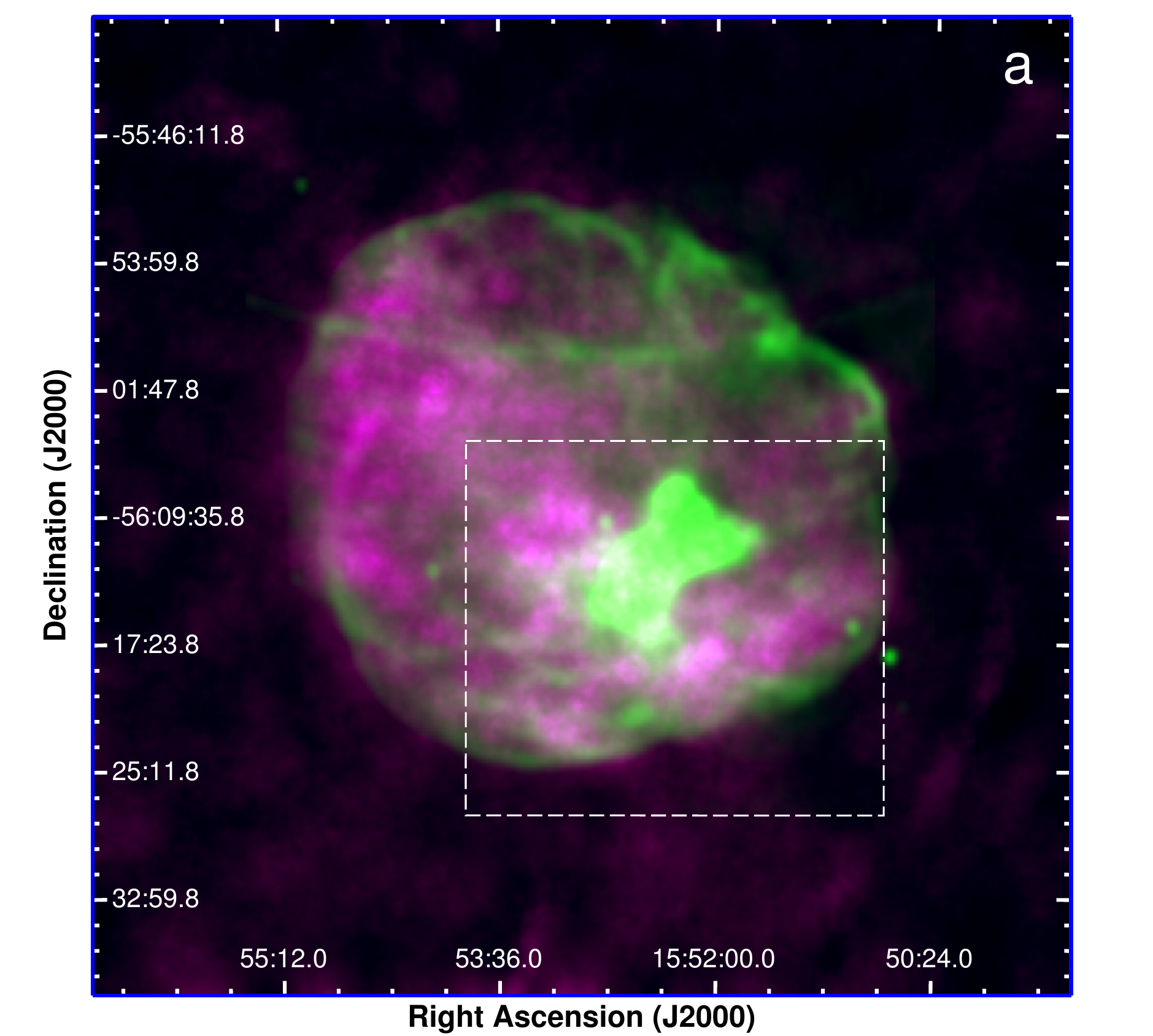}

\epsscale{0.95} \plotone{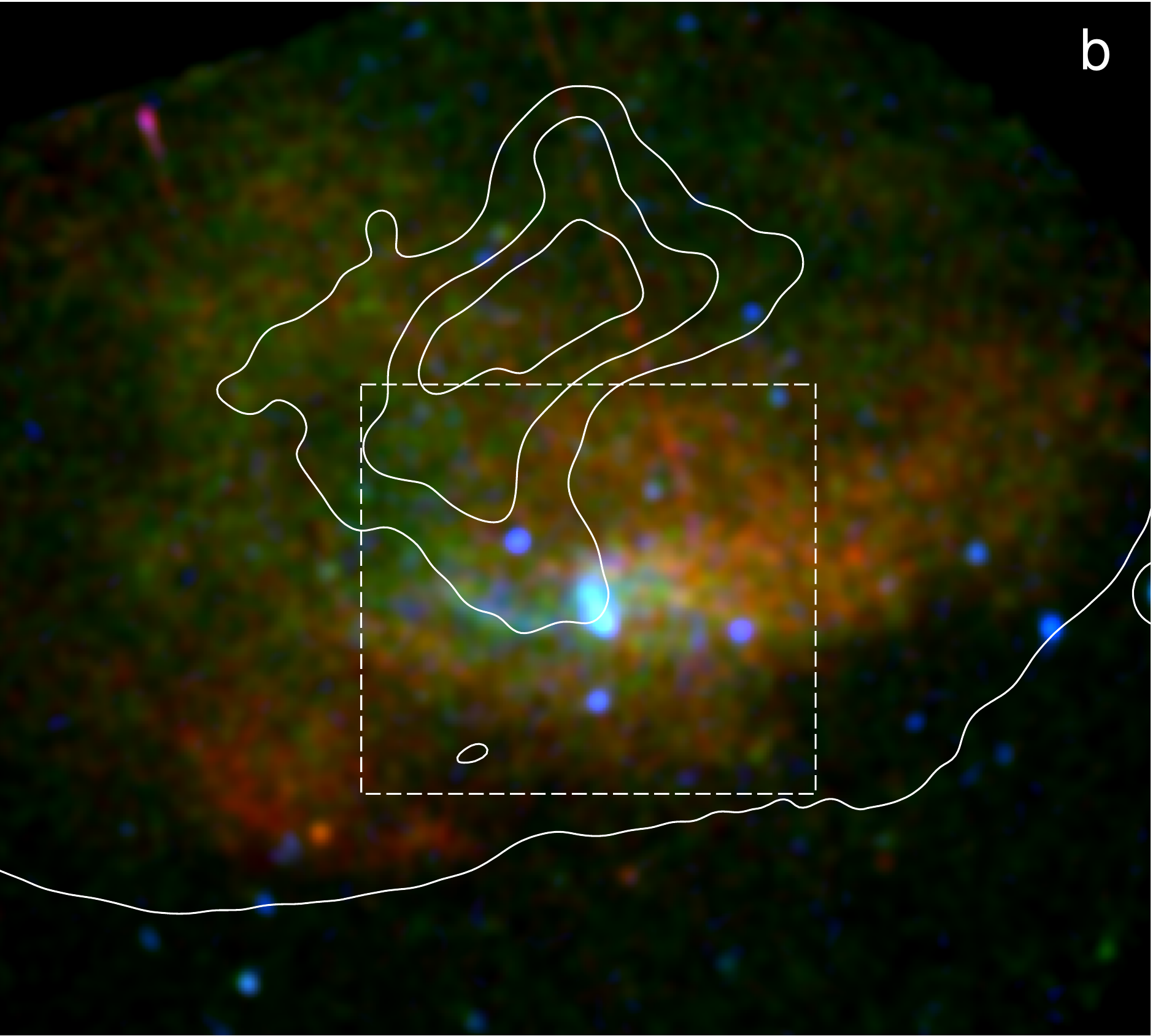}

\epsscale{0.95} \plotone{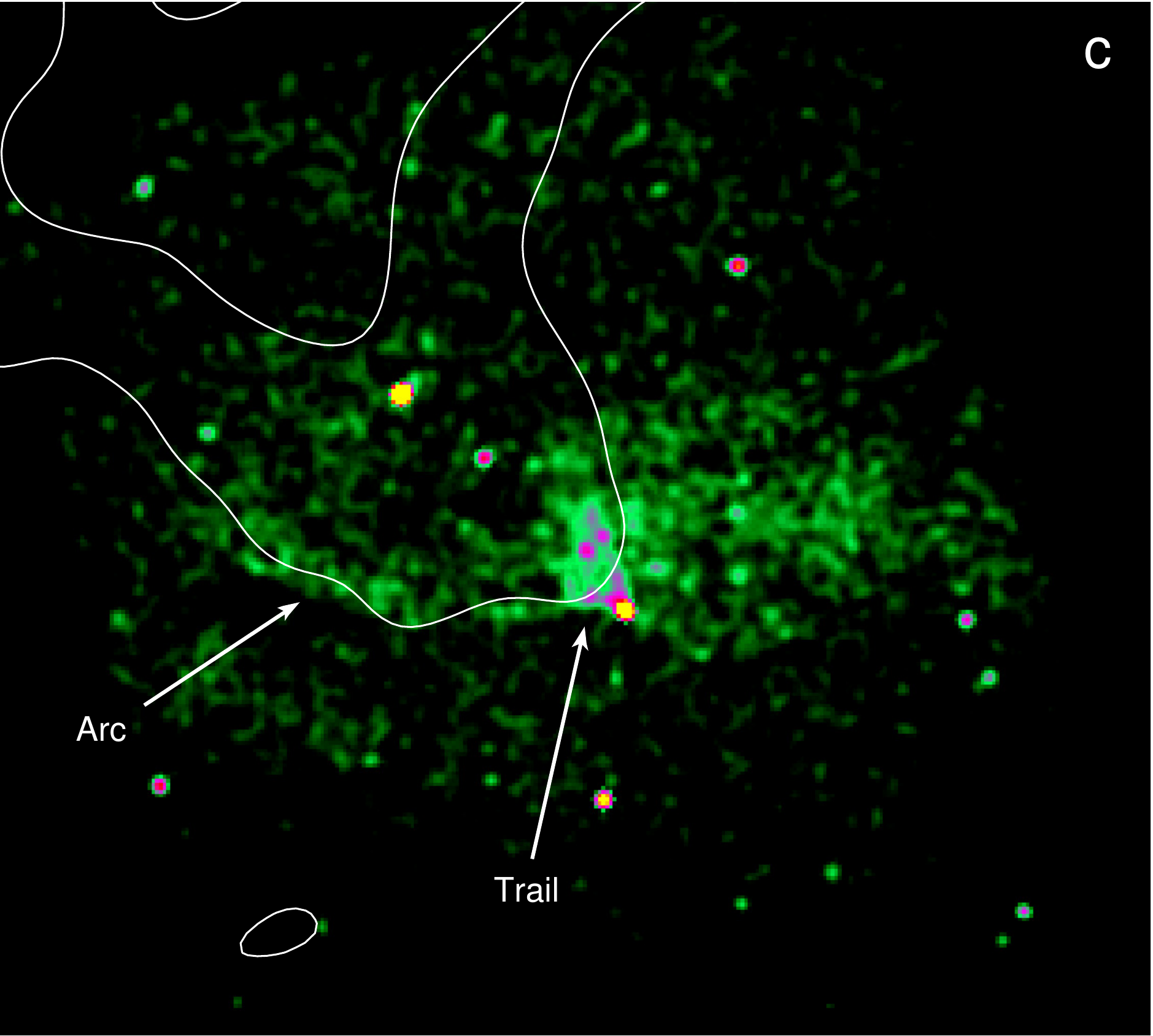}

\caption{\label{images}(a) MOST 843 MHz radio image of MSH~15--5\textit{6} \citep{whi96} in green, and the 0.1-2.4 keV ROSAT X-ray image \citep{kas93} in violet. The region in the white dashed rectangle is shown in panel (b). (b) A combined three color XMM-Newton X-ray image (MOS1/MOS2/pn) with 0.5-1.0 keV emission shown in red, 1.0-3.0 keV image in green, and 3.0-9.0 keV image in blue. MOST radio contours are overlaid in white. The region in the white dashed rectangle is shown in panel (c). (c) Chandra X-ray 0.3-10.0 keV image of the PWN in MSH~15--5\textit{6} resolving the trail behind a point source and an arc of emission tracing the edge of the radio PWN. The MOST radio contours are overlaid in white.}
\end{figure}

\section{OBSERVATIONS AND DATA REDUCTION} \label{obsv}


\subsection{X-ray}

\subsubsection{XMM-Newton}

\textit{XMM-Newton } observations of MSH~15--5\textit{6} were carried out on 2004 August 11 with the MOS and pn cameras, under the observation ID 020427010. The observations were taken in two different exposures for a total exposure time of 56 ks. However, the observations were significantly affected by flares, so after the standard processing with the XMM-SAS software, version 11.0.0,  the resulting exposure times for the  MOS 1, MOS2, and the pn instruments were 29.9 ks, 29.7 ks, and 23.0 ks, respectively. 

\begin{figure*}
\epsscale{1.1} \plottwo{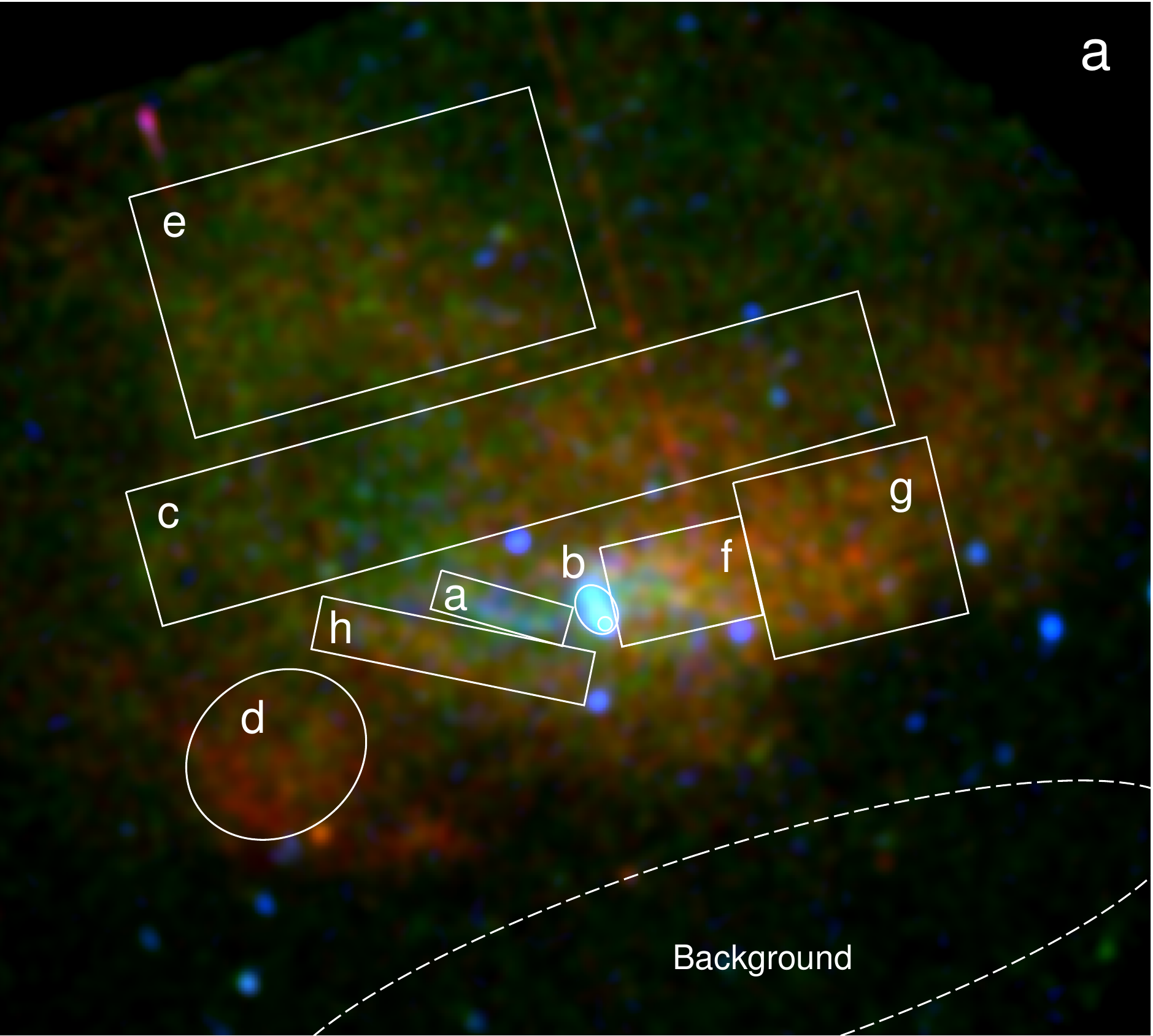}{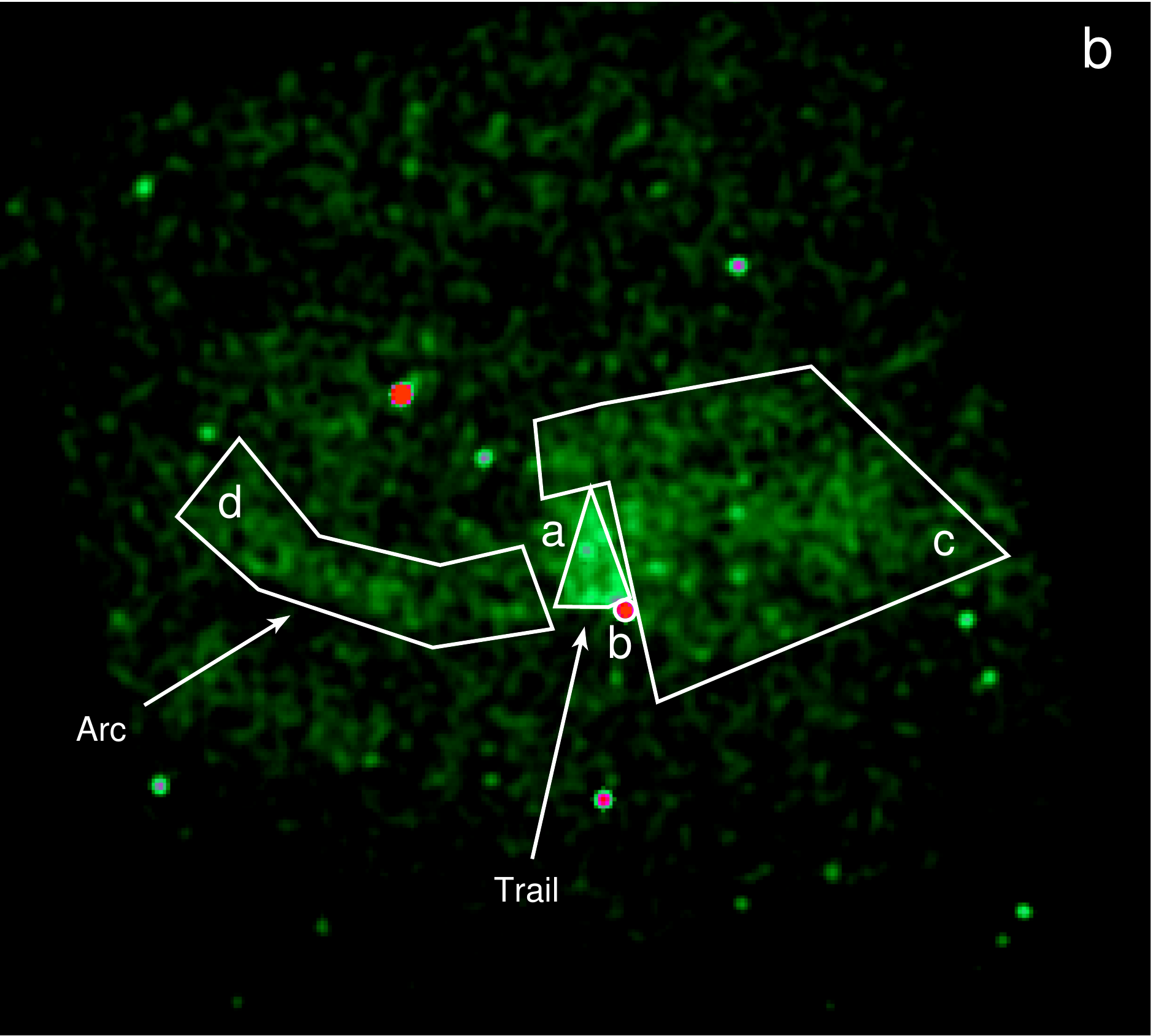}
\caption{\label{regions}(a) A combined three color XMM-Newton X-ray image (MOS1/MOS2/pn) with 0.5-1.0 keV emission shown in red, 1.0-3.0 keV image in green, and 3.0-9.0 keV image in blue. The spectral extraction regions are overlaid in white, along with the dashed background region. (b) Chandra X-ray 0.3-10.0 image with the spectral extraction regions overlaid in white. The figures are on the same spatial scale as those in Figure \ref{images}.}
\end{figure*}

The cleaned data from all three instruments were combined to create exposure-corrected images in three different energy bands using the XMM-SAS \textit{images} script.\footnote{http://xmm.esac.esa.int/sas/current/documentation/threads/ epic\_merging.shtml} The image is shown in Figure \ref{images}. The spatial binning size for the images was set to 5\arcsec. For the purpose of spectral extraction, data from both exposures of the two MOS instruments were merged into a single MOS dataset using the merging recipe from the SAS website (website here). In the same way, the two pn exposures were merged into a single pn dataset. The spectra were then extracted from the merged MOS and pn files using SAS version 11.0.0, using the source and background regions shown in Figure \ref{regions}a, and fitted simultaneously. The spectra were grouped to include at least 30 counts in each bin, background subtracted, and fitted using the CIAO 4.3 Sherpa software. All spectra were fitted with a two-component model consisting of a power-law and the XSVNEI thermal component. The XSPEC model \textit{tbabs} was used for the absorption along the line of sight. We found that the two-component model significantly improved the $\chi^2$ value for all regions.

\subsubsection{Chandra}

MSH~15--5\textit{6} was observed with the \textit{Chandra X-ray Observatory's} Advanced CCD imaging Spectrometer, ACIS-S. Observations were carried out on 2001 August 18 for a total exposure time of 56.8 ks, under the observation ID 1965. The standard cleaning and data reduction were performed using CIAO version 4.3. 
Since the emission from the SNR fills the entire field of view of the ACIS detector, we extracted background spectra from the blank-sky files produced by the ACIS calibration team (http://cxc.harvard.edu/ciao/threads/acisbackground/). A background spectrum extracted from the blank-sky data using the same region was subtracted from each spectrum, and the residual data were binned and fitted using the CIAO 4.3 Sherpa software. The same two-component spectral model as for the \textit{XMM} data was used to fit the \textit{Chandra} spectra.

\subsection{$\gamma$-Ray}

In this work, 51 months of data from {\it Fermi}-LAT (from August 2008 until November 2012) were analyzed. Only events belonging to the Pass 7 V6 {\it Source} class, which reduces the residual background rate as explained in detail in \citet{Abdo2011p7}, were selected for this study. The updated instrument response functions (IRFs) called ``Pass7 version 6'', developed using in-flight data \citep{Rando2009,Abdo2011p7}, were used. Additionally, only events coming from zenith angles smaller than 100$^{\circ}$ were selected so as to reduce the contribution from terrestrial albedo $\gamma$-rays \citep{AbdoPRD}. This study included data from a circular region in the sky centered on the position of MSH 15-56, with radius 20$^\circ$, analyzed using the Fermi Science Tools v9r23p1\footnote{The Science Tools package and related documentation are distributed by the Fermi Science Support Center at http://fermi.gsfc.nasa.gov/ssc}.

The maximum likelihood fitting technique was employed to study the morphological and spectral characteristics of the region \citep{Mattox1996}. The emission models used in {\it gtlike} included a Galactic diffuse component resulting from the interactions of cosmic rays with the ISM and photons (using the mapcube file {\tt gal\_2yearp7v6\_v0\_trim.fits}), and an isotropic one that accounts for the extragalactic diffuse and residual backgrounds (modeled using the {\tt iso\_p7v6source.txt} table). 

In order to understand the spatial characteristics of the $\gamma$-ray emission in the field of MSH 15-56, data in the energy range 2 to 200 GeV, and converted in the {\it front} section, were used. The 68\% containment radius angle for normal incidence {\it front}-selected photons in this energy band is $\leq 0.4^\circ$. Galactic and isotropic backgrounds were modeled and test statistic (TS) maps were constructed using {\it gttsmap} to allow for detection significance estimates, and to best evaluate the position and possible extent of the source.  The test statistic is the logarithmic ratio of the likelihood of a point source being at a given position in a grid, to the likelihood of the model without the additional source, $2log(L_{ps}/L_{null})$. 

The study of the spectral energy distribution (SED) characteristics of the source associated with MSH 15-56, was performed combining events converted in both {\it front} and {\it back} sections, and in the energy range 0.2-204.8 GeV. The lower energy bound was selected to avoid the rapidly changing effective area of the instrument at low energies, and because of the large uncertainty below 0.2 GeV related to the Galactic diffuse model used. {\it gtlike} is used to model the flux at each energy bin and estimate, through the maximum likelihood technique, the best-fit parameters. Background sources from the 24-month {\it Fermi} LAT Second Source Catalog \citep{Abdo2011b}\footnote{ The data for the 1873 sources in the {\it Fermi} LAT Second Source Catalog was made available by the Fermi Science Support Center at http://fermi.gsfc.nasa.gov/ssc/data/access/lat/2yr\_catalog/} were included in the model likelihood fits. The systematic uncertainties of the effective area, for the IRF used, are energy dependent: 10\% at 100 MeV, decreasing to 5\% at 560 MeV, and increasing to 20\% at 10 GeV \citep[and references therein]{Porter2009,Abdo2011b}. As an addition to the statistical uncertainties associated with the likelihood approach, and the systematic errors related to the IRFs, the uncertainty of the underlying Galactic diffuse level was considered. This source of uncertainty was included in the evaluation of the systematic uncertainties by artificially varying the normalization of the Galactic background by $\pm6\%$ from the best-fit value at each energy bin, similarly to the analysis used in \citet{cas10}.

\section{ANALYSIS AND RESULTS} \label{results}

\subsection{Multi-Wavelength Morphology}

SNR MSH~15--5\textit{6} is the prototypical example of a composite SNR. Figure \ref{images}a shows the MOST 843 MHz radio image of MSH~15--5\textit{6} \citep{whi96} in green, and the 0.1-2.4 keV ROSAT X-ray image \citep{kas93} in violet. The radio morphology consists of a symmetric filamentary shell, with a bright PWN component offset to the SW from the geometric center. The shell is roughly circular, and approximately 18\arcmin\ in radius, which corresponds to $\sim$ 21.5 pc for a distance of 4.1 kpc. The shell emission is brightest in the NW, where the X-ray emission appears to be fainter. The shape of the PWN is irregular and suggests that the nebula has likely interacted with the SN reverse shock. It is elongated along the SE/NW direction, with approximate dimensions of 5\farcm6 along the long and 3\farcm4 along the short axis. \citet{dic00} analyzed the high resolution ATCA images of MSH~15--5\textit{6} that revealed more
detailed structures of the PWN, consisting of parallel ridges aligned with the direction of the long axis. The integrated flux densities at 1 GhZ are 114 Jy and 26 Jy for the shell and PWN components, respectively.

The X-ray emission is patchy is appearance and fills the entire SNR shell. The ROSAT soft X-ray emission (0.1-2.4 keV) is seen in violet in Figure \ref{images}a.  There are localized regions of enhanced emission, but they do not correlate with any radio structures. The emission in the whole southeastern half of the remnant and around the PWN is enhanced, but there is no spatial correlation between the diffuse X-ray emission and the radio PWN. \citet{kas93} found that the X-ray emission is mostly thermal, with a temperature of 0.56 keV.

The \textit{XMM-Newton} observations are shown in Figure \ref{images}b, with the 0.5-1.0 keV emission shown in red, 1.0-3.0 keV image in green, and 3.0-9.0 keV image in blue. The MOST radio contours are overlaid in white. The \textit{XMM} field of view includes the region around the radio PWN, and does not cover the entire shell. There is a point source located at the very tip of the radio PWN, and a trail of hard X-ray emission immediately behind the point source, in the direction of the radio PWN. The western emission appears softer, while the emission around the northern part of the PWN appears more green in the three-color image. There is a patch of very soft emission SE of the PWN, near the edge of the radio shell contour. 

The dashed white rectangle in Figure \ref{images}b represents the \textit{Chandra} field of view shown in Figure \ref{images}c. The high-resolution \textit{Chandra} image shows the detailed structure of the X-ray PWN. A point source located at the tip of the radio PWN, at the position  5\fh52\fm12.5\fs , -56$^{\circ}$18$\arcmin$58$\farcs$0. The trail of X-ray emission behind the point source has a bow-shock-like morphology, suggestive of a PWN produced by a high-velocity pulsar. Such cometary morphologies are observed in the Mouse, Geminga, G327.1-1.1, and a number of other PWNe \citep[e.g.][]{kar08}. Similarly, there is an arc of X-ray emission tracing the southern edge of the contour of the radio PWN. Diffuse X-ray emission is present west of the point source, but this emission does not appear to correlate with any radio structures.

\begin{figure*}
\epsscale{1.15} \plottwo{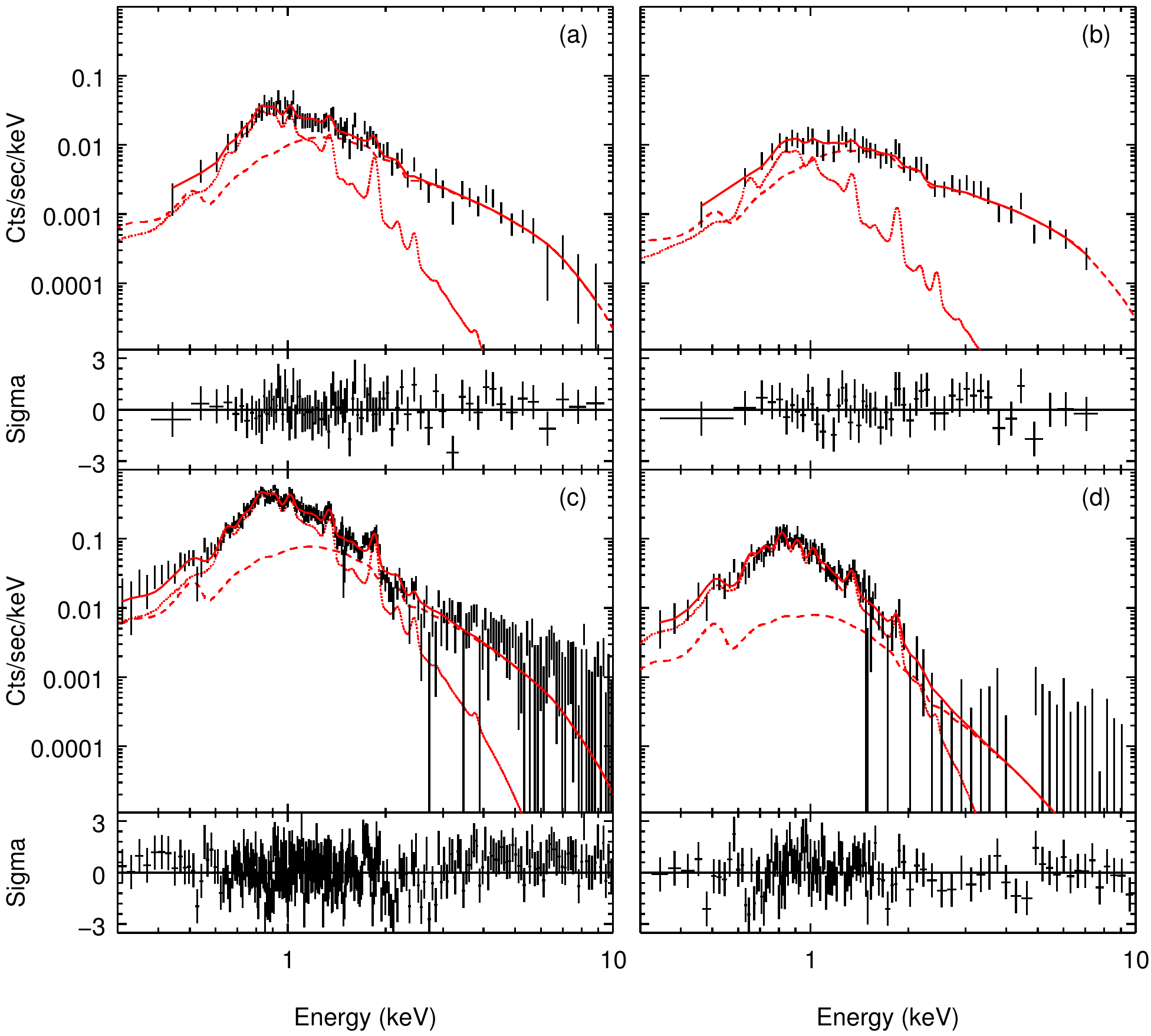}{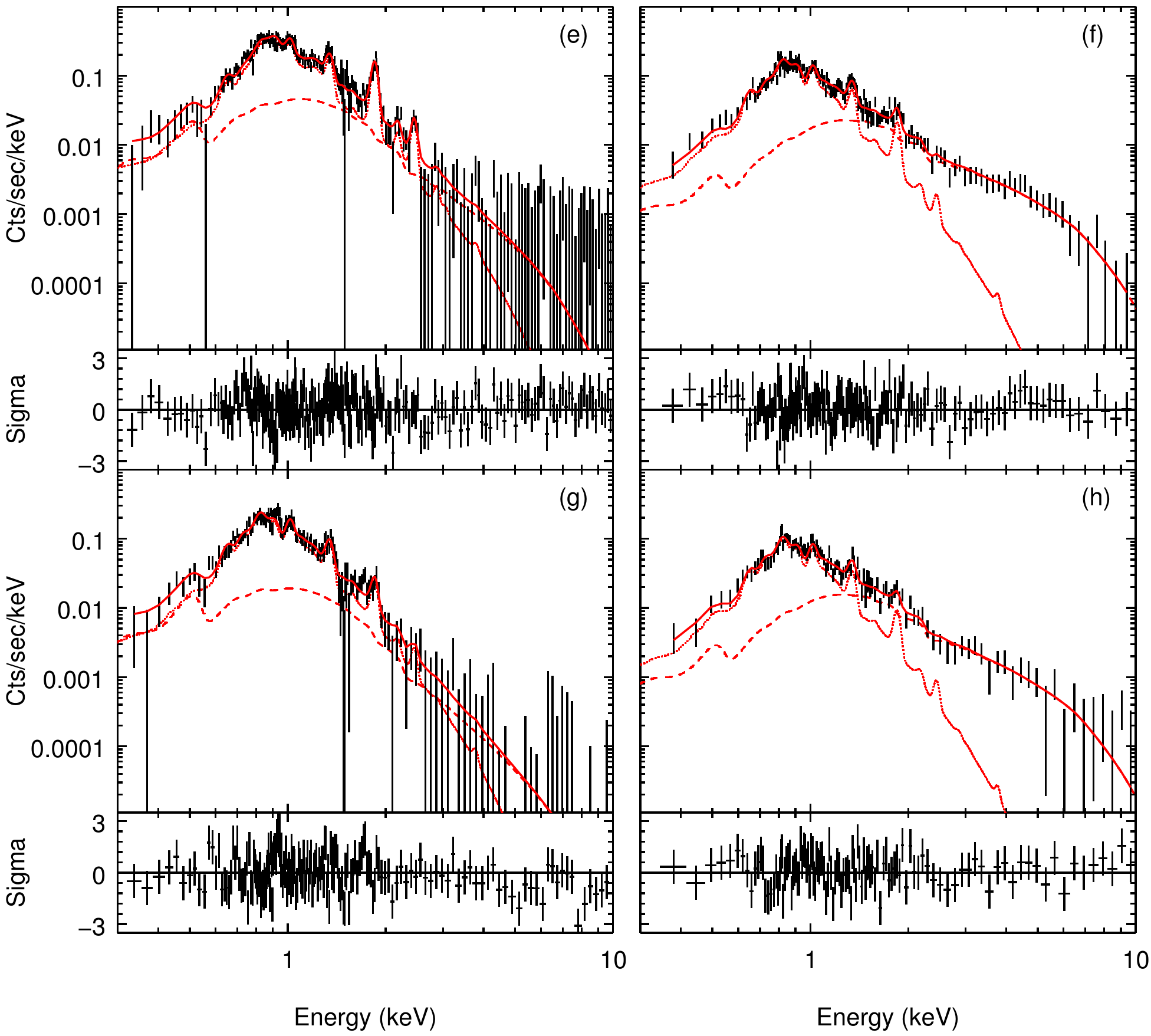}
\caption{\label{xmm_plots} \textit{XMM-Newton} X-ray spectra extracted from regions (a)-(h) in Figure \ref{regions}a. The best-fit models are shown as solid red lines. The individual components are the XSVNEI thermal model (dotted red line) and a power-law model (dashed red line).}
\end{figure*}



\begin{deluxetable*}{lcccc}
\tablecolumns{5} \tablewidth{0pc} \tablecaption{\label{xmmtable}XMM
SPECTRAL FITTING RESULTS} 
\tablehead{ REGION: &
\colhead{a} & \colhead{b} & \colhead{c}  & \colhead{d}} 
\startdata
$\rm N_H(10^{22}\rm cm^{-2})$ & $0.51^{+0.002}_{-0.001}$  & \nodata & \nodata  &  \nodata \\
Photon Index ($\Gamma$) & $2.36^{+0.09}_{-0.25}$ & $2.01^{+0.12}_{-0.11}$  & $3.18^{+0.07}_{-0.07}$  & $4.47^{+0.32}_{-0.30}$ \\
Amplitude &  $(1.1_{-0.3}^{+0.1})\times10^{-4}$ & $(7.5_{-0.8}^{+1.0})\times10^{-5}$ & $(8.2_{-0.4}^{+0.6})\times10^{-4}$ & $(1.4_{-0.2}^{+0.6})\times10^{-4}$ \\
$\rm F_{1} (\rm erg/cm^{2}/s)$ & $5.4\times10^{-13}$ & $4.2\times10^{-13}$ & $4.5\times10^{-12}$ & $1.7\times10^{-12}$ \\
$kT (\rm keV)$ & $0.56_{-0.02}^{+0.20}$ & $0.62_{-0.20}^{+0.30}$ & $0.58_{-0.01}^{+0.01}$  & $0.29_{-0.01}^{+0.04}$   \\
Silicon & $1.6_{-1.1}^{+0.6}$ & $4.5_{-2.3}^{+3.8}$ & $1.5_{-0.1}^{+0.1}$ & $1.7_{-0.8}^{+0.8}$ \\
Sulfur & $6.6_{-5.4}^{+4.1}$ & $0.0_{-...}^{+13}$ & $1.3_{-0.8}^{+0.7}$ & $1.0_{-...}^{+4.9}$ \\
$\rm \tau (s\:cm^{-3})$ & $(2.3_{-1.5}^{+0.8})\times10^{11}$ & $(1.1_{-0.7}^{+2.6})\times10^{11}$ &  $(1.8_{-0.1}^{+0.1})\times10^{11}$ & $(3.5_{-3.4}^{+...})\times10^{13}$  \\
Normalization &  $(1.5_{-0.5}^{+0.1})\times10^{-4}$ & $(3.9_{-1.3}^{+2.7})\times10^{-5}$ & $(2.0_{-0.1}^{+0.1})\times10^{-3}$ & $(2.2_{-0.5}^{+0.1})\times10^{-3}$ \\
$\rm F_{2} (\rm erg/cm^{2}/s)$ & $4.7\times10^{-13}$ & $1.8\times10^{-13}$ & $7.2\times10^{-12}$ & $4.3\times10^{-12}$ \\

Reduced $\chi^2$ Statistic & 0.7 & 0.7 & 1.1 & 1.0 \\
\hline
\hline
\\
REGION: & e & f & g & h \\
\hline
$\rm N_H(10^{22}\rm cm^{-2})$ & \nodata  & \nodata &  \nodata  &  \nodata \\
Photon Index ($\Gamma$) & $3.80^{+0.17}_{-0.19}$ & $2.30^{+0.05}_{-0.26}$  & $4.54^{+1.98}_{-0.20}$  & $2.82^{+0.08}_{-0.07}$ \\
Amplitude &  $(8.4_{-1.3}^{+1.1})\times10^{-4}$ & $(1.7_{-0.5}^{+0.1})\times10^{-4}$ & $(2.5_{-1.5}^{+0.2})\times10^{-4}$ & $(4.4_{-0.3}^{+0.4})\times10^{-4}$ \\
$\rm F_{1} (\rm erg/cm^{2}/s)$ & $6.4\times10^{-12}$ & $8.6\times10^{-13}$ & $3.3\times10^{-12}$ & $7.0\times10^{-13}$ \\
$kT (\rm keV)$ & $0.66_{-0.07}^{+0.01}$ & $0.51_{-0.01}^{+0.10}$ & $0.51_{-0.01}^{+0.14}$  & $0.54_{-0.01}^{+0.01}$   \\
Silicon & $2.9_{-0.2}^{+0.7}$ & $1.2_{-0.6}^{+0.2}$ & $0.9_{-0.5}^{+0.2}$ & $1.3_{-0.6}^{+0.4}$ \\
Sulfur & $3.7_{-0.2}^{+0.2}$ & $1.0_{-...}^{+1.4}$ & $0.8_{-...}^{+1.0}$ & $0.0_{-...}^{+2.1}$ \\
$\rm \tau (s\:cm^{-3})$ & $(1.6_{-0.1}^{+0.1})\times10^{11}$ & $(1.9_{-0.8}^{+0.2})\times10^{11}$ &  $(2.1_{-1.1}^{+0.2})\times10^{11}$ & $(2.0_{-0.1}^{+0.1})\times10^{11}$  \\
Normalization &  $(2.2_{-0.1}^{+0.5})\times10^{-3}$ & $(7.0_{-2.0}^{+0.1})\times10^{-4}$ & $(1.2_{-0.4}^{+0.1})\times10^{-3}$ & $(1.4_{-0.1}^{+0.1})\times10^{-3}$ \\
$\rm F_{2}  (\rm erg/cm^{2}/s)$ & $8.7\times10^{-12}$ & $2.5\times10^{-12}$ & $4.1\times10^{-12}$ & $1.7\times10^{-12}$ \\
Reduced $\chi^2$ Statistic & 0.9 & 0.8 & 1.1 & 0.8 \\

\enddata
\tablecomments{The listed uncertainties are 1-sigma statistical uncertainties. The apertures used in the spectral extractions are shown in Figure \ref{regions}a, and the spectra and fits are shown in Figure \ref{xmm_plots}. The fluxes were calculated in the 0.3-10 keV band. The normalization of the thermal models is equal to $10^{-14}n_e n_H V / 4\pi d^2 \: \rm cm^{-5}$, where $V$ is the volume of the emitting region and $d$ is the distance to the SNR.}
\end{deluxetable*}


\begin{figure*}
\epsscale{0.9} \plotone{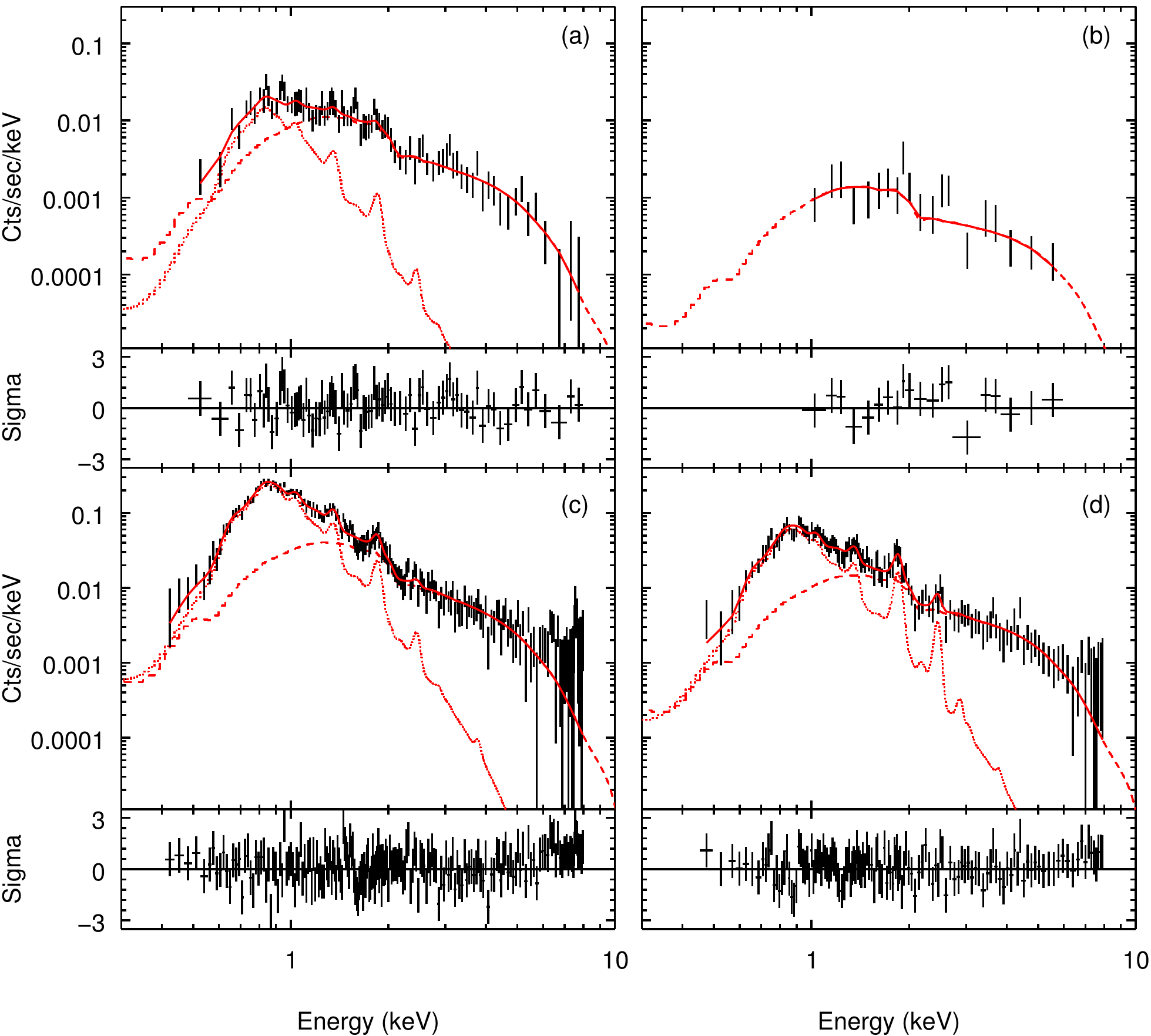}
\caption{\label{chandra_plots} Chandra X-ray spectra extracted from regions in Figure \ref{regions}b. The best-fit models are shown as solid red lines. The individual components are the XSVNEI thermal model (dotted red line) and a power-law model (dashed red line).}
\end{figure*}



\begin{deluxetable*}{lcccc}
\tablecolumns{5} \tablewidth{0pc} \tablecaption{\label{chandratable}CHANDRA
SPECTRAL FITTING RESULTS} 
\tablehead{ REGION: &
\colhead{a} & \colhead{b} & \colhead{c}  & \colhead{d}} 
\startdata
$\rm N_H(10^{22}\rm cm^{-2})$ & 0.51  & \nodata & \nodata  &  \nodata \\
Photon Index ($\Gamma$) & $1.84^{+0.13}_{-0.11}$ & $1.41^{+0.24}_{-0.24}$  & $2.05^{+0.11}_{-0.08}$  & $1.60^{+0.12}_{-0.13}$ \\
Amplitude &  $(5.2_{-0.6}^{+0.7})\times10^{-5}$ & $(5.6_{-1.2}^{+1.3})\times10^{-6}$ & $(2.1_{-0.2}^{+0.2})\times10^{-4}$ & $(6.6_{-0.9}^{+0.9})\times10^{-5}$ \\
$\rm F_{1} (\rm erg/cm^{2}/s)$ & $3.2\times10^{-13}$ & $5.2\times10^{-14}$ & $1.1\times10^{-12}$ & $5.0\times10^{-13}$ \\
$kT (\rm keV)$ & $0.49_{-0.13}^{+0.10}$ & \nodata & $0.56_{-0.02}^{+0.01}$  & $0.62_{-0.04}^{+0.04}$   \\
Silicon & $3.9_{-2.6}^{+4.6}$  & \nodata & $1.0_{-0.3}^{+0.2}$ & $2.8_{-0.6}^{+0.6}$ \\
Sulfur & $0.0_{-...}^{+13}$  & \nodata & $1.0_{-...}^{+1.4}$ & $5.6_{-2.9}^{+2.9}$ \\
$\rm \tau (s\:cm^{-3})$ & $(2.5_{-1.3}^{+...})\times10^{11}$ & \nodata &  $(1.8_{-0.2}^{+0.1})\times10^{11}$ & $(1.7_{-0.4}^{+0.4})\times10^{11}$  \\
Normalization &  $(4.0_{-1.2}^{+0.2})\times10^{-5}$ &\nodata & $(5.9_{-0.3}^{+0.2})\times10^{-4}$ & $(1.4_{-0.1}^{+0.1})\times10^{-4}$ \\
$\rm F_{2} (\rm erg/cm^{2}/s)$ & $1.3\times10^{-13}$ & \nodata & $2.1\times10^{-12}$ & $5.4\times10^{-13}$ \\

Reduced $\chi^2$ Statistic & 0.7 & 0.9 & 0.9 & 0.6 \\
\enddata
\tablecomments{The listed uncertainties are 1-sigma statistical uncertainties. 
The apertures used in the spectral extractions are shown in Figure \ref{regions}b, and the spectra and fits are shown in Figure \ref{chandra_plots}. 
The fluxes were calculated in the 0.3-10 keV band. The normalization of the thermal models is equal to $10^{-14}n_e n_H V / 4\pi d^2 \: \rm cm^{-5}$, where $V$ is the volume of the emitting region and $d$ is the distance to the SNR.}
\end{deluxetable*}


\subsection{X-Ray Spectroscopy}

X-ray spectra were extracted from the \textit{XMM} and \textit{Chandra} data from the regions shown in Figure \ref{regions}. For the \textit{XMM} data, we selected eight different regions that correspond either to specific structures identified in the \textit{Chandra} image, or to regions that show differences in colors in the three-color \textit{XMM} image. The selected spectral regions in Figure \ref{regions}a are the following; (a) the arc structure that traces the edge of the radio PWN, (b) the cometary X-ray trail behind the point source, (c) the central region of the diffuse X-ray emission, (d) the softest X-ray patch that appears red in the \textit{XMM} three-color image, (e) the northern region that appears most green in the \textit{XMM} three-color image, (f, g) the diffuse X-ray emission east of the PWN, and (h) the larger, more diffuse arc of X-ray emission visible in the \textit{Chandra} image. A region outside of the SNR's radio contours was selected for the background exctraction, and this region is indicated by the dashed white ellipse in Figure \ref{regions}a. The spectral extraction regions used for the \textit{Chandra} data are shown in Figure \ref{regions}b. The regions correspond to distinct structures; (a) the cometary X-ray trail, (b) the X-ray point-like source at the tip of the radio PWN, (c) the emission east of the PWN, and (d) the arc tracing the edge of the radio PWN. The background for the \chandra data was extracted from the blank-sky data, as described in Section \ref{obsv}. The extracted \xmm and \chandra spectra are shown in Figures \ref{xmm_plots} and \ref{chandra_plots}, respectively.

We first attempted to fit the X-ray spectra with a single XSVNEI model, but found that a high-energy excess required an additional power-law component for most regions. We therefore fitted all the spectra with a two-component model, a power-law plus a thermal XSVNEI model, using the XSPEC absorption model TBABS. The point source spectrum was the only one that was fitted with a power-law component only. The absorption along the line of sight was linked and simultaneously fitted for all the \xmm regions. The best fit value of $\rm N_H=0.51\times10^{22}\: cm^{-3}$ was then used as a frozen parameter for the fitting of the \chandra spectra. The abundances of silicon and sulfur were clearly enhanced in region (e) of Figure \ref{regions}a, the northernmost region in the image, so we thawed the silicon and sulfur abundances for all the fits in order to determine if the there is a possible enhancement. The best fit parameters for the \xmm and \chandra fits are listed in Tables \ref{xmmtable} and \ref{chandratable}.

Based on the best fit parameters, we make the following observations:


(1) All regions besides the soft X-ray patch (region (d) of Figure \ref{regions}a) are fitted with approximately the same thermal temperature of around 0.5-0.6 keV and the same ionization timescale of approximately $2\times10^{11}\: \rm s\:cm^{-3}$.

(2) Region (e) of Figure \ref{regions}a shows enhanced abundances of silicon and sulfur that are up to 3 or 4 times the solar value, providing clear evidence that the emission in this region comes from SN ejecta. The arc feature in Figure \ref{regions}b also shows enhanced abundances, consistent with the interpretation that this emission comes from SN ejecta that have been swept up by the PWN. The abundances in other regions are not well constrained, so we cannot discount the possibility that the emission from other regions also comes from SN ejecta.

(3) Region (d) is the only region for which the best-fit
parameters of the thermal component significantly
differ from the others. The best-fit temperature is
0.3 keV and the ionization timescale is significantly
longer, on the order of $10^{13}\: \rm s\:cm^{-3}$.

(4) The photon index steepens with distance from the point source, varying from $\Gamma=2.3$ in the cometary PWN structure to $\Gamma=4.5$ in the outer regions.  Similar steepening with distance from the pulsar has been observed in other PWNe, including 3C 58, G21.5-0.9, G327.1-1.1, and G54.1+0.3 \citep{sla00,sla04,tem09,tem10}. The steepening can be explained qualitatively by synchrotron
aging of particles as they travel from the pulsar towards the outer nebula \citep{ken84}. The large photon index of $\Gamma>3.5$ that is seen in the outermost regions labeled (e), (d), and (g) in Figure \ref{regions}a is somewhat steeper than expected. We suggest that an underlying soft X-ray component from swept-up ISM is contributing to the X-ray spectrum and steepening the apparent photon indices of the power-law component (see Section \ref{ejecta}). Even though this additional soft X-ray component cannot be constrained by the current data, our simulations show that the presence of such a component with a temperature of $\sim$ 0.3 keV and a surface brightness three times lower than in region (d) would raise the temperature of the presumed ejecta component, and flatten the global photon index of the power-law component from $\Gamma=3.0\pm0.1$ to $\Gamma=2.4\pm0.1$. 

The total luminosity of the X-ray emission is $L_X\rm(0.3-10 \:keV)=4.0\times10^{34}\: erg\:s^{-1}$. Using a value of $\dot{E} = 4 \times
10^{36} {\rm\ erg} {\rm\ s}^{-1}$ (see Section \ref{gray}), this results in an X-ray radiation efficiency of 0.01. While this value is somewhat higher than average, it falls well within the scatter in the $L_X/\dot{E}$ relationship observed for other pulsar \citep[e.g.][]{pos02}.

(5) The spectrum of the point source is fitted with a power-law with a photon index of $\Gamma=1.4\pm0.2$, consistent with the value expected for a pulsar. The unabsorbed flux $F_X\rm(0.3-10\: keV)=5.2\times10^{-14}\: erg\:cm^{-2}\:s^{-1}$, translating into a luminosity of $L_X\rm(0.3-10 \:keV)=1.0\times10^{32}\: erg\:s^{-1}$ (for a distance of 4.1 kpc), a value that is in the range of other young pulsars \citep{kar08}. 


The interpretation of the best-fit parameters is discussed in Section \ref{thermal}.

\subsection{$\gamma$-Ray Emission}

A $\gamma$-ray source is coincident with the position of the PWN with centroid $(\alpha_{2000},\delta_{2000}=15^{h}52^{m}45^{s}, -56^{\circ}11'46.7'')$, with $95\%$ confidence radius $=3'.6$. The X-ray point source is located near the edge, while the PWN is enclosed entirely within the $95\%$ confidence circle. The origin of the $\gamma$-ray emission cannot be constrained based on the location of centroid alone, since both the pulsar, the PWN, and the SNR itself are likely candidates. We discuss these possible scenarios for the origin of the $\gamma$-ray emission in Section \ref{gray}.

\begin{figure}
\epsscale{1.1} \plotone{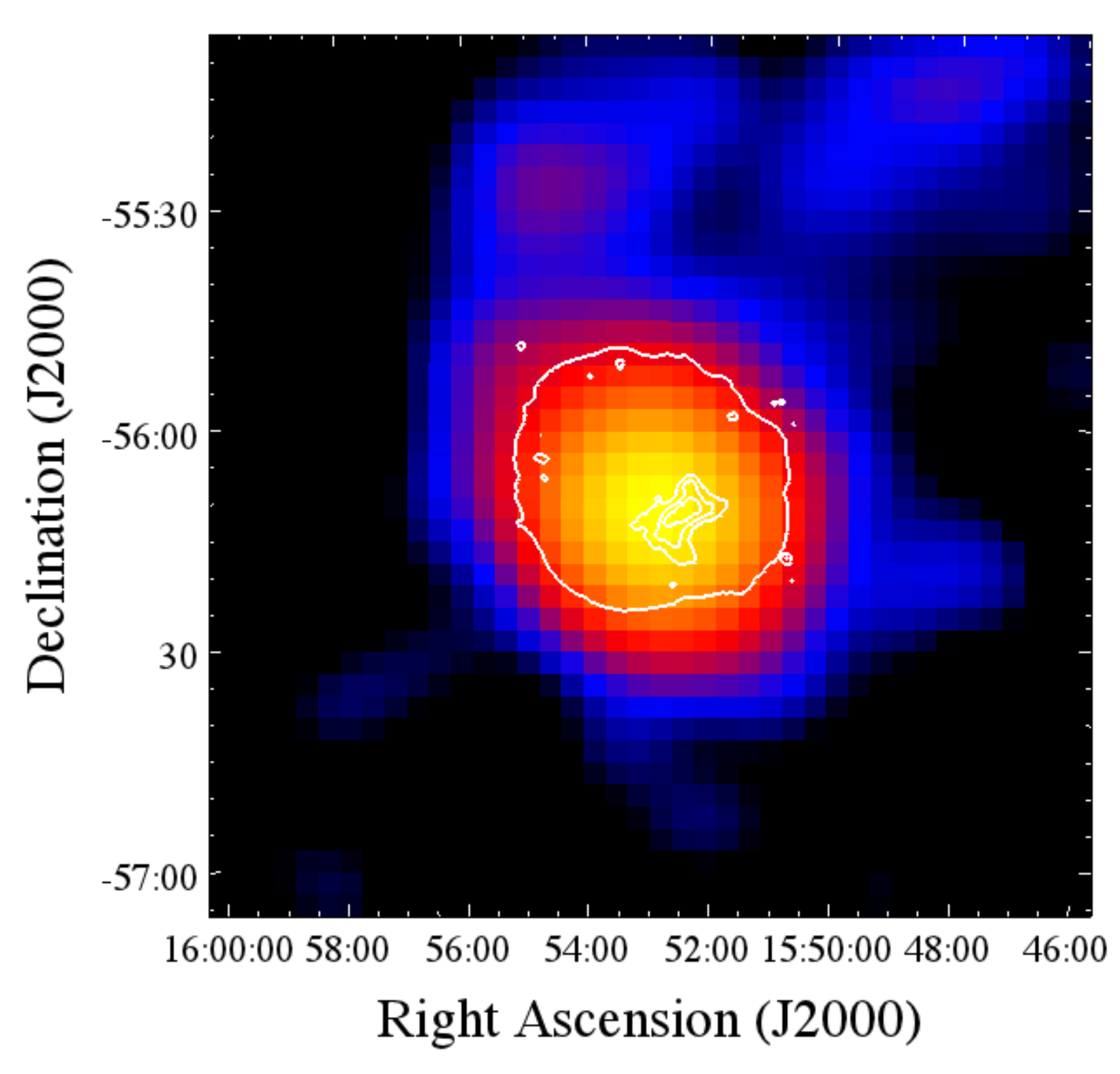}
\caption{\label{ts_map}\textit{Fermi}-LAT $\gamma$-ray emission from MSH~15--5\textit{6} with the MOST radio contours overlaid in white. The \textit{Fermi} source appears to be centered at the position of the disrupted PWN. }
\end{figure}

\begin{figure}
\epsscale{1.2} \plotone{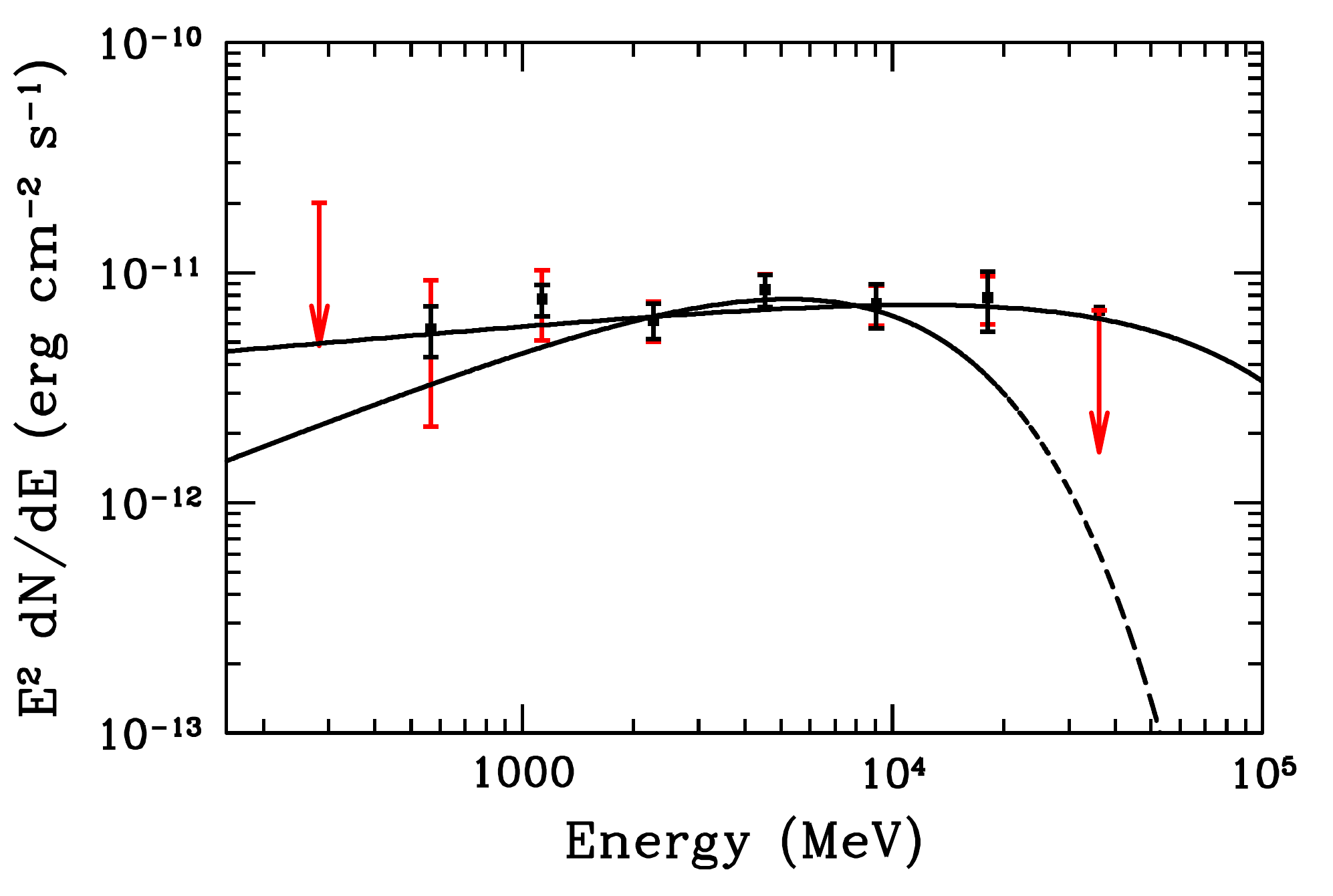}
\caption{\label{gammaspec}\textit{Fermi}-LAT spectrum of emission from MSH 15Ð56. Also shown are model fits 
for a power law with an exponential cutoff. The solid curve is derived 
by fixing the flux in the highest energy bin to be equal to the derived 
upper limit. For the dashed curve, the flux in the bin was fixed at a 
value ten times below the upper limit.}
\end{figure}

The significance of the detection, obtained from the evaluation of the peak of the TS map in Figure \ref{ts_map}, is $\sim 14.2\sigma$.
The \textit{Fermi} LAT spectrum from MSH 15-56 is shown in Figure \ref{gammaspec}, along with 
fits to a power law model with an exponential cutoff. The solid curve 
corresponds to the model with the flux in the highest energy fixed at 
the value of the derived upper limit shown in the plot, while the dashed 
curve corresponds to the same model with the flux in this bin fixed at a 
factor of 10 below the derived upper limit. Assuming a distance of 4.1 kpc, the range in fit parameters 
from these two fits is $E_{\rm cut} \approx 8.2-43$~GeV, $\Gamma = 1.4-1.9$, and $F(>100 {\rm\ MeV}) = (1.5-3.3) \times 10^{-8}{\rm\ 
photons\ cm}^{-2}{\rm\ s}^{-1}$. The associated luminosity is $L_{\gamma} (> 100 \rm\ MeV) = (2.6-5.5) \times 10^{35} \rm
erg \rm\ s^{-1}$. The ratio $L_{\gamma}/\dot{E}$ in this case is somewhat higher than for other PWNe, which may in part be explained by the increase in $\gamma$-ray luminosity following the collision between the PWN and the reverse shock \citep{gel09}.

\section{Origin of the Thermal X-Ray Emission}\label{thermal}

The thermal emission from the northernmost region in the \xmm image, region labeled (e) in Figure \ref{regions}a, clearly shows silicon and sulfur abundances that are significantly higher than solar abundances. This is compelling evidence that the emission from this region originates at least in part from SN ejecta. The best-fit temperature and ionization timescale in this region are $0.66^{+0.01}_{-0.07}$ keV and $1.6\times10^{11}\:\rm s\: cm^{-3}$, respectively. The abundances are also enhanced in the arc structure in Figure \ref{images}c, with similar values for the best-fit temperature and ionization timescale. While the abundances in the other regions are not well constrained,
the temperatures and ionization timescales are within uncertainties for all regions besides region (d) of Figure \ref{regions}a. The smaller patch of soft X-ray emission in region (d) of Figure \ref{regions}a has a best-fit temperature of $0.30^{+0.04}_{-0.01}$ keV, significantly lower than for other regions, and a lower limit on the ionization timescale of $10^{13}\rm \: s \: cm^{-3}$. 

Based on these spectral properties, it is plausible that much of the X-ray thermal emission that we observe with \chandra and \xmm (all regions besides region (d) in Figure \ref{regions}a) is dominated
by ejecta that have been heated by the passage of the SN reverse shock, or swept-up or mixed-in with the PWN. The enhanced X-ray emission at the edges of the PWN likely arises from SN ejecta that have been swept up and heated by the nebula.  The emission in region (d) of Figure \ref{regions}a may originate from the swept-up ISM in the SNR shell. However, there are some discrepancies associated with this interpretation that will be discussed further in Section \ref{ejecta}.

\section{Derived SNR Properties}\label{properties}

We used the best-fit temperature and normalization of the thermal component in region (d) of Figure \ref{regions}a, and a distance of 4.1 kpc  to calculate the SNR radius $R$ and the shock velocity $v_s$, under
the assumption of electron-ion temperature equilibration, which
appears appropriate for the inferred  electron temperature of the
shocked ISM component \citep{gha07}.
 
\begin{equation}
R(pc) = 21\: D_{4.1}
\end{equation}

\begin{equation}
v_s(km \: s^{-1}) = 500 \:T^{1/2}_{0.29\:keV}
\end{equation}

Assuming the \citet{sed59} model for the evolution of the SNR in the Sedov-Taylor stage, we calculate the initial ISM density $n_0$, SNR age $t$, explosion energy $E$, and the swept-up ISM mass $M$. We find

\begin{equation}
n_0(cm^{-3}) = 0.1 \:D^{-1/2}_{4.1}
\end{equation}

\begin{equation}
t(kyr) = 16.5\: D_{4.1} T^{-1/2}_{0.29\:keV}
\end{equation}

\begin{equation}
E(10^{51}\:erg) = 1.0 \:D^{5/2}_{4.1}
\end{equation}

\begin{equation}
M(M_{\odot}) = 100 \:D^{3}_{4.1}
\end{equation}

The initial ISM density $n_0$ was calculated assuming that $n_0=n_e/4$, where $n_e$ is derived from the normalization of the thermal spectrum equal to $10^{-14}n_e n_H V / 4\pi d^2 \: \rm cm^{-5}$, and a spherical emitting volume with a radius of 2\farcm7 for region (d) in Figure \ref{regions}a. The derived parameters are not unreasonable for a remnant in the Sedov-Taylor stage of its evolution. We note that the model assumes a uniform ISM density, and a single average thermal temperature for the entire swept-up shell. The uncertainties in the SNR distance and non-uniform ambient density introduce unquantifiable uncertainties in the derived parameter values. Based on the derived SNR age, we also estimate the transverse velocity of the presumed pulsar. The displacement of the point source from the geometric center of the shell is approximately 5\farcm8, which translates into a transverse velocity of 410 $\rm km\: s^{-1}$, within the range of typical pulsar velocities.

\begin{deluxetable}{lc}
\tablecolumns{2} \tablecaption{\label{sedov}DERIVED SNR PROPERTIES} 
\tablewidth{17pc}
\tablehead{\colhead{Property} & \colhead{Value}} 
\startdata
D($\rm kpc$) & 4.1 \\
R($\rm pc$) & 21 \\
$v_s(\rm km\:s^{-1})$& 500 \\
t($\rm yr$) & 16,500 \\
n$_0$($\rm cm^{-3}$) & 0.1 \\
$E_{51}$ ($10^{51}\rm erg$) & 1.0 \\
$M(\rm M_{\odot}^{}$) & 100 \\
$v_{psr}(\rm km\:s^{-1})$ & 410 \\
\enddata
\end{deluxetable}

\section{Presence of SN Ejecta}\label{ejecta}

The enhanced abundances and similar ionization timescales of all 
regions besides region (d) in Figure \ref{regions}a suggest that some of the 
observed X-ray thermal emission originates from SN ejecta that have
been either heated by the SN reverse shock, or the PWN. The observed 
thermal emission is most likely a mixture of emission from the swept-up 
material and ejecta.  We can place an upper limit on the ejecta mass by 
assuming that all of this thermal emission arises from the ejecta 
(assuming all of the ejecta have been shocked). Based on the derived 
spectral properties of the thermal emission, we used two different 
methods to estimate the mass of this component. The two estimates come
from two different derivations of the electron density $n_e$; one from 
the best-fit ionization timescale and the derived SNR age, and the 
other from the best-fit normalization of the thermal spectrum.

Using the age of 16,500 yr and the ionization timescale of $2\times10^{11}\:\rm s\: cm^{-3}$ for the presumed SN ejecta component, we find the electron density to be $n_{e}\rm=0.38\:cm^{-3}$. For a spherical emitting volume with a radius of 6\farcm0 (7 pc for a distance of 4.1 kpc) that approximately encompasses the X-ray emission visible in the \xmm image, we find a total hydrogen mass of $M_{H}=11f\rm \: M_{\odot}$, where $f$ is the volume filling factor.
If we assume that the ejecta are dominated by oxygen, the ejecta density $n_{ej} \approx n_e/8$, and the mass of each atom is assumed to be $16m_p$, where $m_p$ is the mass of a proton. This leads to a total SN ejecta mass of $M_{ej}=25f\rm \: M_{\odot}$.

We can also derive the electron density $n_e$ using the best-fit normalization of the thermal spectrum for the same sized region. The fitted normalization is equal $0.01 \rm cm^{-5}$, which we set equal to $10^{-14}n_e n_H V / 4\pi d^2 \: \rm cm^{-5}$, and find $n_{e}f^{-1/2}\rm=0.24 \: cm^{-3}$ and  $M_{H}=7f^{1/2}\rm \: M_{\odot}$. We note that the best-fit parameters for this global spectrum are 0.55 keV and $2\times10^{11}\:\rm s\: cm^{-3}$, for the thermal temperature and ionization timescale, respectively, and $\Gamma=3.1$ for the photon index of the non-thermal component that accounts for $\sim$33\% of total unabsorbed flux.

In order to estimate the total mass of ejecta dominated by oxygen rather than hydrogen, and correctly account for the dependence of the normalization of the Bremsstrahlung continuum on the square of the atomic number ($Z^2$), we use the following method. We fit the global X-ray spectrum to an absorbed power-law model, fixed 
to the best-fit parameters listed above, plus a bremsstrahlung model 
with a temperature fixed to 0.55 keV. To reduce the contribution from 
thermal line emission, we limited the spectral fit to energies above 2 
keV. We find a best-fit normalization of $\sim 0.01 {\rm\ cm}^{-5}$ 
which, assuming emission from a metal-rich plasma dominated by oxygen, 
and further assuming $n_e \approx 8 n_i$, yields an ejecta density 
$n_{ej} \sim 0.02f^{-1/2} {\rm\ cm}^{-3}$ and a total ejecta mass of $M_{ej} 
\approx 9.6f^{1/2} \rm M_\odot$.

\begin{figure}
\epsscale{1.0}\plotone{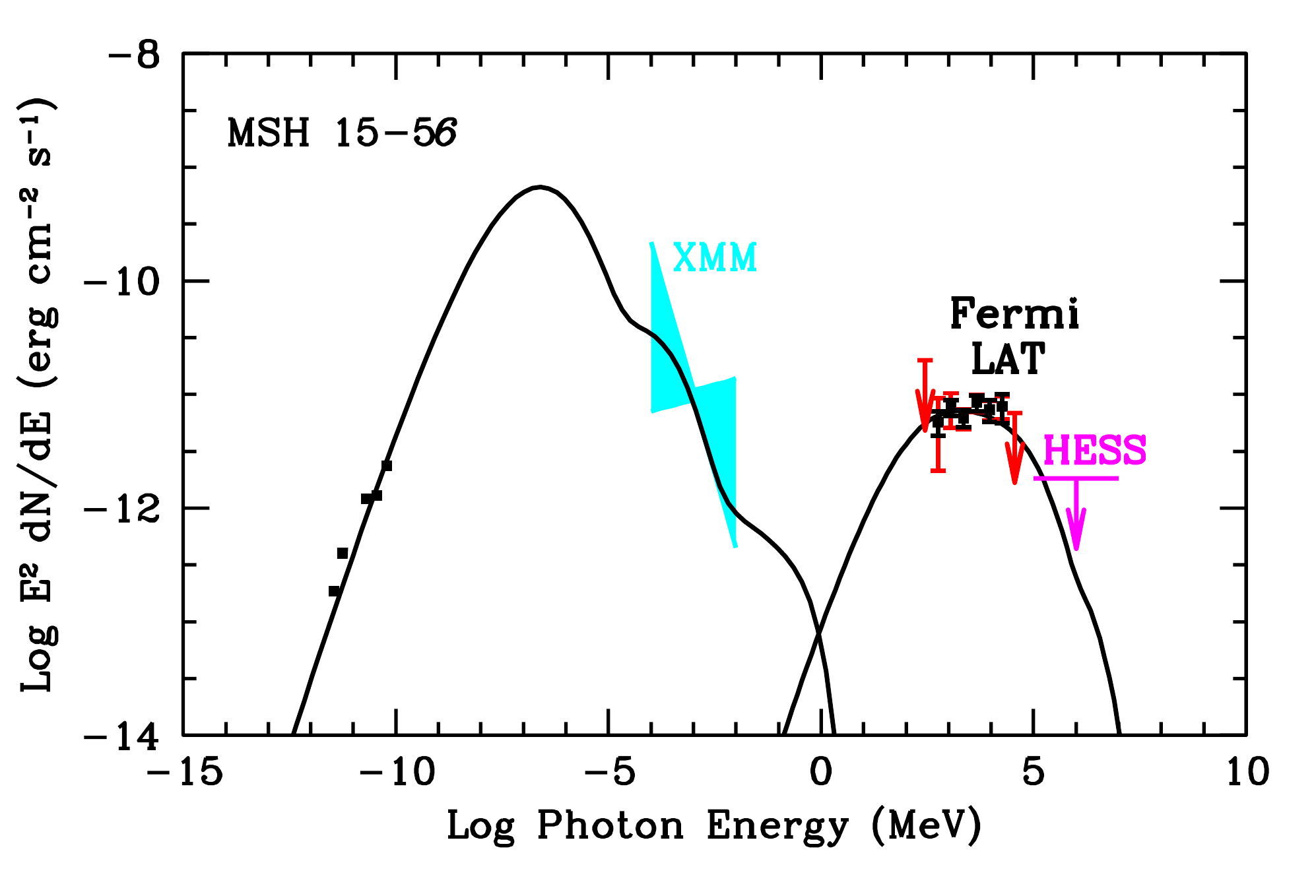}
\epsscale{1.0} \plotone{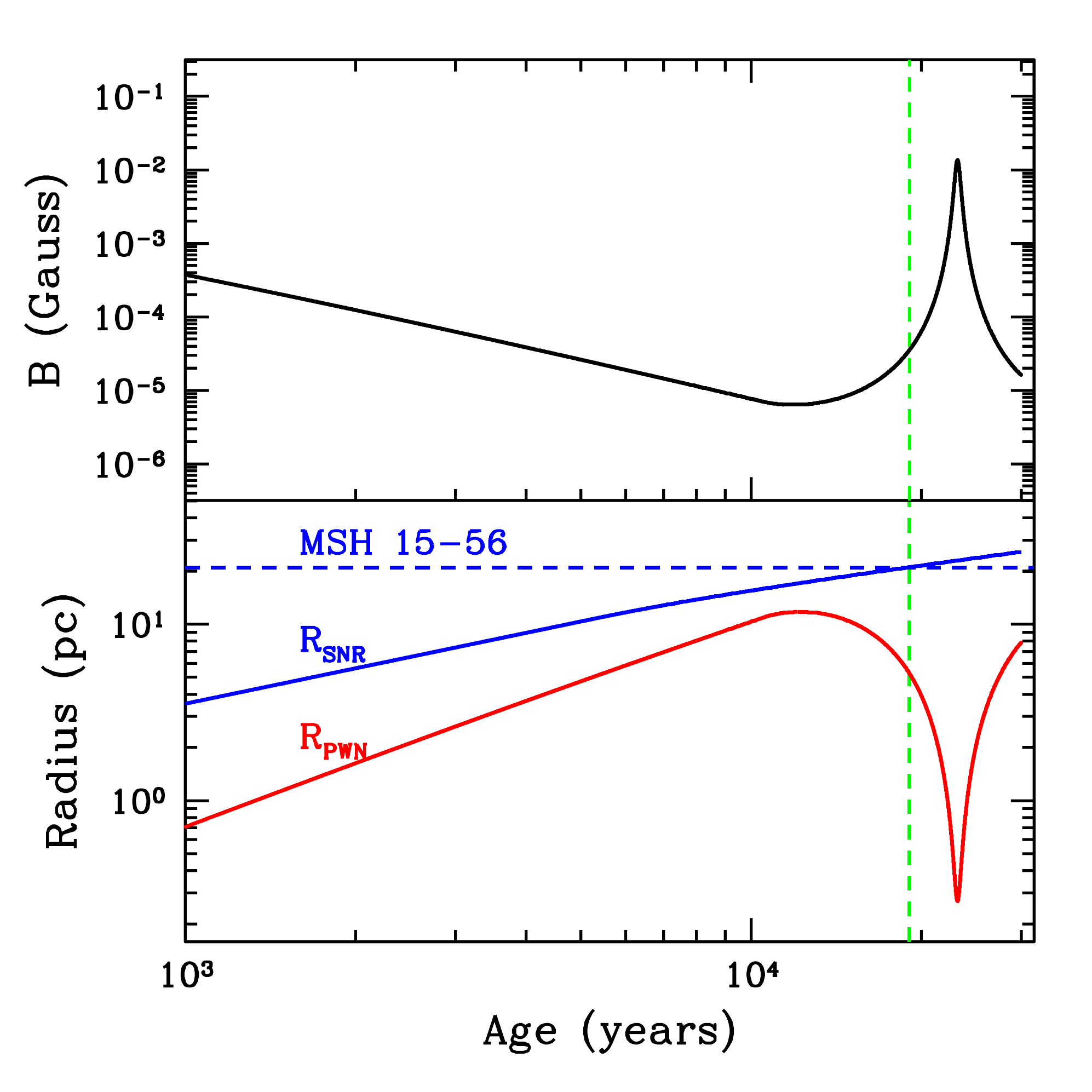}
\caption{\label{models}Top: Model for emission from the evolved PWN in \msh\
(see Section \ref{gray}) along with observed $\gamma$-ray emission (and upper
limit), radio emission and X-ray flux and spectral index from the
PWN. Bottom: Time evolution of the PWN magnetic field (upper panel)
and SNR and PWN radii (lower panel) for \msh. The observed SNR radius
is indicated by the horizontal dashed line, and the age at which
this radius is reached in the model is indicated by the vertical
dashed line.}
\end{figure}

\begin{figure}
\epsscale{1.1}\plotone{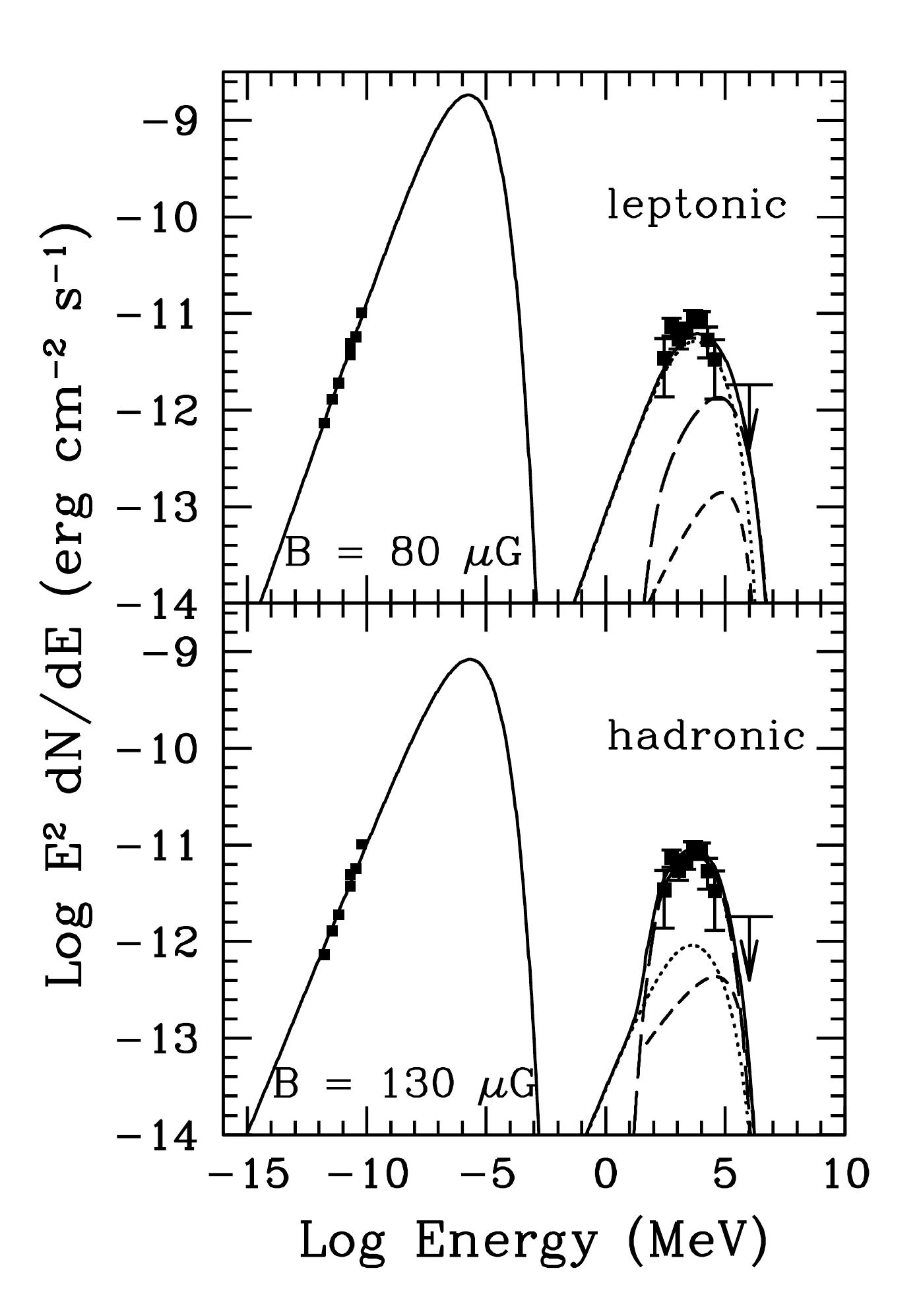}
\caption{\label{models2}Models for the broadband emission from \msh\ assuming the
$\gamma$-ray emission originates from the SNR, dominated either by
leptonic process (upper panel) or hadronic processes (lower panel). 
The dotted curves correspond to inverse Compton emission, short dashed
curves represent nonthermal bremsstrahlung, and long dashed curves
show the emission from $pi^0$ decay.}
\end{figure}

The estimated mass for pure metal ejecta is too high for existing progenitor models
\citep{woo02}, especially considering the fact that the emitting volume used in the calculation is only a fraction of the entire SNR volume.  
The enhanced abundances are a clear indication that ejecta are present, but the contribution of this component may have been overestimated by the fit. A lower filling factor for the ejecta would reduce the calculated mass.
Another possible explanation is that an additional thermal component is contributing to the thermal spectrum in all regions, but is not accounted for by our fits due to the poor statistics of the existing spectra. This component may originate from the swept-up ISM that contaminates the spectrum along the line of sight. On the other hand, multiple components may be present if  there is macroscopic mixing of metal-rich ejecta with hydrogen-rich gas. 

Not accounting for the presence of two separate thermal components may be artificially increasing the normalization of the metal-rich ejecta component that in turn leads to a higher inferred ejecta mass. In order to test this possibility, we attempted to fit the global spectrum of the SNR with a three-component model, a power-law, plus two thermal components for the SN ejecta and a possible soft ISM component (with a temperature fixed to 0.3 keV). We find that the presence of this additional ISM component could raise the temperature of the ejecta component to approximately 0.9 keV, and lower its normalization by a factor of $\sim$5, which would in turn lower the estimated ejecta mass to $\sim4\rm \: M_{\odot}$.
Another possibility is that the ejecta are present only in some regions where we see enhanced abundances, but that most of the emission in other regions does not originate from ejecta. In this case, it would be difficult to explain why both the ejecta and swept-up ISM components have the same temperature and ionization timescale. Deeper observations of the SNR are required to better constrain the spectral parameters and ejecta mass, and confirm the origin of the X-ray thermal emission.

\section{Origin of the $\gamma$-Ray Emission}\label{gray}

We identified \textit{Fermi}-LAT emission from 2FGL J1552.8-5609 coincident
with \msh\ (Figure \ref{ts_map}). The LAT spectrum is shown in Figure \ref{models}, along
with the radio spectrum \citep{dic00}, the
derived flux and spectral index range for the nonthermal X-ray
spectrum from XMM data, and an upper limit to the TeV emission based
on HESS non-detection in an accumulated $\sim 40$ hours of observations
\citep{hop08}. The evolution of the composite system was modeled
following \citet{gel09}, using an injected pulsar spectrum
with $\dot{E}_0 = 3 \times 10^{38} {\rm\ erg} {\rm\ s}^{-1}$,
using a power law spectrum with $\alpha_e = 2.2$ accompanied by a
Maxwellian component with $E_{peak} = 32$~GeV comprising 98\% of
the total electron energy. We assumed that 90\% of the injected
wind was in the form of electron/positron pairs, and that the PWN
expanded into an ejecta core of $25 M_\odot$ driving the SNR into
an ambient density $n_0 = 0.3 {\rm\ cm}^{-3}$, roughly consistent
with that suggested by the X-ray data.

The resulting evolution of the SNR and PWN radii, along with the
PWN magnetic field, are shown in Figure \ref{models}. The model reaches the
observed SNR radius at an age of $1.9 \time 10^4$~yr, by which point the
SNR reverse shock has already begun significantly compressing the
PWN. The PWN magnetic field is $B_{PWN} = 34 \mu{\rm G}$.

The radio and nonthermal X-ray emission (Figure \ref{models}) is well-described
by synchrotron emission from the evolved PWN wind, with the radio
emission dominated by the Maxwellian electron component, and the
X-ray data showing the effects of burn-off of the highest energy
particles due to the increased magnetic field during the PWN
compression phase. The $\gamma$-ray emission is produced through
Inverse Compton scattering off of a three-component field composed
of photons from the CMB, starlight ($T = 3500$~K), and local dust
($T = 50$~K), with equal assumed energy densities. The current
spin-down power of the pulsar in the model is $\dot{E} = 4 \times
10^{36} {\rm\ erg} {\rm\ s}^{-1}$ with a characteristic
age $\tau_c = 2.1 \times 10^4$~yr and an initial spin period $P_0
= 98$~ms. We note the limitations of the 1-D model applied here,
particularly given the observed asymmetric morphology of the PWN.

Young pulsars have been shown to form a significant component of
the Galactic population of GeV $\gamma$-ray sources \citep{abd10}, 
and it is distinctly possible that the \textit{Fermi}
LAT emission observed from \msh\ originates directly from the pulsar
powering the observed radio/X-ray PWN. It is important to note that
the observed cutoff for the power law spectrum is considerably
higher than that observed for the pulsar population, suggesting
that such an association is unlikely, but pulsation searches --
both in the gamma-ray band and in other band -- are of particular
importance to address this scenario. A search of the ATNF pulsar
catalog\footnote{http://www.atnf.csiro.au/research/pulsar/psrcat/}
\citep{man05} reveals no pulsars within
more than 2.5 degrees of the \textit{Fermi} LAT source whose spin-down power
is larger than the observed $\gamma$-ray luminosity at the associated
pulsar positions.

An alternative possibility is that the observed \textit{Fermi} LAT emission
originates from relativistic particles accelerated by the SNR itself.
We find that models in which the emission is produced predominantly
by either electrons, through Inverse Compton scattering of the same
photon field described above for the PWN scenario, or by protons,
through $p-p$ collisions and subsequent $\pi^0$ decay, can both
reproduce the observed broadband emission (Figure \ref{models2}), though not
without some difficulties. Both models require magnetic fields that
are much higher than expected for the compressed ISM, and also low
cutoff energies for the particle spectra. For the leptonic scenario,
$E_{e,cut} \approx 0.5$~TeV and $B_{SNR} \approx 80 \mu{\rm G}$, with
nearly $4 \times 10^{50}$~erg in relativistic particles assuming a
typical electron-proton ratio $K_{e-p} = 10^{-2}$. The low maximum
energy for the electrons is consistent with the high magnetic field,
but the field strength itself is quite large.

For the hadronic scenario, if we assume 40\% of the available SNR
energy (assumed to be $10^{51}$~erg) goes into relativistic particles,
we require a mean density $\bar{n} = 2 {\rm\ cm}^{-2}$ and $K_{e-p}
< 5 \times 10^{-3}$ to reproduce the $\gamma$-ray spectrum. The
associated magnetic field required to produce the observed radio
emission is $B_{SNR} = 130 \mu{\rm G}$. The proton spectrum requires
an exponential cutoff with $E_{p,cut} \approx 0.5$~TeV, which is
surprisingly low given that protons do not suffer significant
radiative losses. The required density is much larger than that
inferred from X-ray measurements. Such results have been observed
for SNRs interacting with molecular clouds \citep{cas10},
possibly suggesting that those SNRs evolve through a clumpy ISM,
with a low interclump density mitigating the expansion and producing
the X-rays, but with the relativistic protons that produce the
$\gamma$-rays interacting with both the high density clumps and the
lower density interclump gas. Such a scenario could also apply to
\msh, although there has been no reported evidence of any interactions
with dense clouds for this SNR.


\section{CONCLUSIONS}
 
\textit{XMM-Newton} and \chandra X-ray observations of the composite SNR MSH~15--5\textit{6} reveal complex structures that provide evidence for mixing of the SN ejecta with PWN material following a reverse shock interaction. An X-ray point source is located at the southern tip of the radio PWN, with a trail of hard X-ray emission immediately behind the point source, indicative of a newly formed X-ray PWN. Enhanced X-ray emission traces the edges of the radio PWN and may arise from SN ejecta that have been swept up and heated by the nebula.

The X-ray spectra are well described by a two-component model, a non-thermal power-law plus a thermal, non-equilbrium ionization model model with $\rm N_H=0.5\times10^{22}\: cm^{-3}$. The thermal emission shows evidence for
both SN ejecta and the swept-up ISM shell. Select regions clearly show Si and S abundances a factor of $\sim$3 higher than solar, and a best-fit temperature and ionization timescale of $0.66^{+0.01}_{-0.07}$ keV and $(1.6\pm0.2)\times10^{11}\:\rm s\: cm^{-3}$, respectively. 
While the abundances are not constrained in all regions, the similarities in temperature and ionization timescale suggest that much of the X-ray emission originates from SN ejecta that have either been heated by the reverse shock or swept up by the PWN. We suggest that there also may be a significant underlying ISM component contributing to the spectrum that cannot be constrained by the current data. Deeper X-ray observations are required to measure the relative contribution of SN ejecta and the swept-up ISM.  We do find one region along the southern edge of the SNR shell that has a lower temperature of $0.30^{+0.04}_{-0.01}$ keV, and appears to be in ionization equilibrium. Assuming the Sedov model, we derive an SNR age of 16,500 yr. 

The origin of the observed $\gamma$-ray emission remains somewhat
unclear. The most likely scenario appears to be that it originates
from the evolved PWN. However, emission directly from the pulsar
powering the observed nebula, as well as emission from the SNR 
itself if the mean magnetic field is very high -- and perhaps
if the remnant is expanding into a clumpy ISM, may also provide
plausible solutions. Sensitive pulsation searches from the pulsar
are of considerable importance to address this issue, as are deeper
X-ray and molecular line measurements of the SNR in order to better
assess the properties of the ambient medium.

\acknowledgments



\end{document}